\documentclass[11pt]{article}
\pdfoutput=1

\usepackage[utf8]{inputenc}
\usepackage[T1]{fontenc}
\usepackage{amsmath,amssymb}
\usepackage{graphicx}
\usepackage{booktabs}
\usepackage[hidelinks]{hyperref}
\usepackage[round]{natbib}
\usepackage{geometry}
\usepackage{placeins}
\graphicspath{{./}}

\title{Where does the criticality live?\\
Early-warning signals are event-heterogeneous across\\
seven crypto-perpetual liquidation cascades}

\author{Ramon Marc Garcia Seuma\thanks{Correspondence: reymon.devs@gmail.com}}
\date{\today}

\begin{document}
\maketitle

\begin{abstract}
Do crypto perpetual-futures crashes carry a reproducible early-warning fingerprint
of a critical transition, and in \emph{which} state variable? We study seven major
BTC liquidation cascades (2022--2025, including the record \$19B event of
10 October 2025) using minute-level price and 5-minute leverage/order-flow data.
On detrended residuals we compute rolling variance and lag-1 autocorrelation and
test their pre-cascade trend with the Kendall-$\tau$ statistic, sweeping 39
analysis configurations per variable per event. \emph{No variable is
event-invariant.} Price carries the critical-slowing-down signature in five of
seven events but is silent in exactly the two sudden-news (tariff) shocks,
suggesting a two-type structure: endogenous-buildup versus exogenous-shock
cascades. The October 2025 event---whose in-sample analysis suggests the
signature lives in leverage rather than price---turns out to be the outlier, not
the rule. The one regularity surviving all events with data is a
\emph{compression} of taker order-flow variance, which passes a 300-onset
placebo test (Fisher-combined $p\approx 5\times10^{-6}$) but is a
population-level precursor, not a per-event alarm. Single-event
critical-slowing-down claims in crypto derivatives are therefore fragile by
construction. We argue the pattern of failures is itself diagnostic: slowing
down is absent exactly where the destabilising mechanism is most abrupt, as one
would expect if these cascades are discontinuous, shock-driven transitions
rather than critical ones.
\end{abstract}

\section{Introduction}
\label{sec:intro}

On 10 October 2025 the market for crypto perpetual futures produced the largest
liquidation cascade on record: upwards of \$19\,billion of leveraged positions
were force-closed across roughly 1.6 million accounts within a few hours, an
order of magnitude beyond any prior episode \citep{amberdata2025oct}. The
proximate trigger was exogenous---an unscheduled announcement of 100\% tariffs
on Chinese imports \citep{cnbc2025tariff}---but the machinery that converted a few
per cent of price movement into a market-wide cascade is endogenous, and it is
written into the exchange rulebook. Positions are leveraged; an adverse move
crosses maintenance-margin thresholds; the exchange's liquidation engine emits
forced, price-insensitive sell orders; their market impact pushes the price
further in the same direction, into the next thresholds. This is the positive
feedback by which leverage manufactures fat tails and clustered volatility
\citep{thurner2012leverage}, and by which market liquidity and funding liquidity
become mutually reinforcing \citep{brunnermeier2009market}.

Cascades of this kind admit two readings, and they differ in a way that ought to
be empirically decidable. On the first, the crash is an exogenous shock: news
arrives, the price adjusts, and leverage merely amplifies a move that would have
occurred anyway. On the second, the market has been drifting toward an
instability for some time and the news is simply the perturbation that finds it
there---the crash is the crossing of a critical point, preceded by a slow
approach that a sufficiently attentive observer could in principle detect.
Econophysics has long pursued the second reading, modelling crashes as critical
points with precursory signatures \citep{johansen2000crashes}. Which reading is
right is not an academic matter: only under the second does a pre-crash state
variable exist to be measured.

The theory of early-warning signals makes the second reading operational. As a
dynamical system approaches a fold bifurcation, the leading eigenvalue of its
linearisation tends to zero and perturbations decay ever more slowly---critical
slowing down---which surfaces in the observable statistics as rising variance
and rising lag-1 autocorrelation
\citep{scheffer2009early,dakos2012methods,scheffer2012anticipating}. The
framework's founding review is, however, candid about the distance between the
theory and complex reality: ``the question therefore remains of whether highly
complex real systems \ldots\ will show the theoretically expected early-warning
signals'' \citep{scheffer2009early}. That same review reports ``empirical
successes in finding early-warning signals of transitions in systems for which
we have a relatively poor understanding of the mechanisms that drive the
dynamics, such as the human brain and financial markets''. The financial half of
this claim has since been contested from within the same tradition:
\citet{guttal2016lack} find no critical slowing down ahead of historical
financial meltdowns and conclude that they are not critical transitions, and
\citet{diks2019critical} reach a compatible verdict.

\citet{scheffer2009early} also names the methodological problem that we take as
our point of departure. Observing that indicator trends ``depend on parameter
choices in filtering'', it argues that it ``would be useful to build a set of
reliable statistical procedures to test whether an increase in autocorrelation,
for example, is significant''. We take that instruction literally, and it
organises this paper. Every claim below is evaluated across a 39-configuration
sweep of the detrending and rolling-window choices; a hypothesis formed on one
event is tested out of sample on events that did not generate it; and the single
regularity that survives is checked against an explicit placebo null of ordinary
market windows. This discipline is not decoration. It is what turns our own most
attractive in-sample finding into a negative result.

Crypto perpetual futures are an unusually good laboratory for the question, and
not because they crash often. Early-warning theory concerns the \emph{state
variable} of a dynamical system, and for a leveraged market it is not obvious
that the state variable is the price. The mechanism sketched above implicates
leverage and order flow, and perpetual exchanges publish precisely those objects,
for free, at five-minute resolution: aggregate open interest, the long/short
positioning of ordinary accounts and of the largest traders, and the aggressor
side of every trade. We therefore do not merely ask whether early-warning signals
precede a cascade. We ask \emph{in which variable} they live, and whether that
answer is stable from one cascade to the next. It is not.

This paper makes deliberately narrow, falsifiable claims and stress-tests them
across events rather than within one. Our contribution is fourfold:
\begin{enumerate}
  \item a reproducible pipeline reconstructing seven major cascades (2022--2025)
        from public data (Section~\ref{sec:data});
  \item evidence that the most-cited early-warning signal, rising variance, is
        \emph{non-specific} throughout (Section~\ref{sec:results});
  \item a demonstration that crash signatures are \emph{event-heterogeneous}:
        the in-sample October 2025 pattern (autocorrelation in leverage/flow,
        absent in price) inverts out of sample, and across seven events no
        variable is invariant---with a candidate two-type structure
        (endogenous-buildup vs.\ exogenous-shock)
        (Sections~\ref{subsec:oos}--\ref{subsec:distribution});
  \item a placebo-tested positive result: pre-cascade \emph{compression} of
        taker order-flow variance is the sole population-level regularity
        (Section~\ref{subsec:placebo}).
\end{enumerate}

Section~\ref{sec:data} describes the data and Section~\ref{sec:methods} the
estimator and the robustness protocol. Section~\ref{sec:results} presents the
results in the order in which they were obtained---October 2025 in sample,
August 2024 out of sample, then the seven-event panel and the placebo
test---because that order is the argument. Sections~\ref{sec:discussion} and
\ref{sec:conclusion} ask what the failure of single-variable early warning
implies, and where it points.

\section{Data}
\label{sec:data}

All data are public and the pipeline is scripted (repository
\texttt{critical-phenomena-in-crypto-perps}). We use Binance USD-margined (UM)
perpetual futures for BTCUSDT around seven major cascades: May 2022 (LUNA/UST),
November 2022 (FTX), August 2024 (yen-carry unwind), December 2024,
February 2025 (tariff scare), April 2025 (``Liberation Day''), and October 2025
(the record collateral-depeg cascade). The narrative below develops the October
2025 event in sample, tests it out of sample on August 2024, then widens to the
full panel. For each event, a $\sim$2-month window:
\begin{itemize}
  \item \textbf{Price}: 1-minute klines ($\approx$88{,}000 bars per event window)
        from \texttt{data.binance.vision} (the REST API is geo-restricted for our host).
  \item \textbf{Leverage/order-flow}: 5-minute \emph{metrics} dumps
        ($\approx$17{,}600 rows per window): open interest (USD), top-trader
        long/short ratio, global long/short
        account ratio, and taker buy/sell volume ratio.
  \item \textbf{Context}: a daily derivatives panel (2022--2025) of open interest,
        funding and long/short liquidation notionals.
\end{itemize}
\paragraph{What the public data cannot show.}
Two limitations are structural rather than incidental, and they shape the whole
analysis. First, Binance's per-event \texttt{liquidationSnapshot} dumps have been
discontinued and are no longer served for any date, so intraday liquidation
microstructure---the very object on which the cascade mechanism acts---is
unavailable at this resolution without a commercial data source. Daily
long/short liquidation notionals support the anatomy in
Section~\ref{sec:results}; for the intraday analysis we necessarily observe the
\emph{consequences} of forced liquidation (open interest, aggressor flow) rather
than liquidations themselves. Second, funding settles every eight hours, far too
coarsely to serve as an intraday early-warning variable; it enters only as
context. Our leverage variables are therefore proxies for the liquidation state,
not measurements of it, and every claim below should be read with that
substitution in mind. Section~\ref{sec:conclusion} returns to it.

\section{Methods}
\label{sec:methods}

\paragraph{State variable and detrending.}
For a series $x_t$ we form a stationary residual by subtracting a trailing moving
average of window $w_d$,
\begin{equation}
  r_t = x_t - \frac{1}{w_d}\sum_{k=0}^{w_d-1} x_{t-k}.
\end{equation}
For price we take $x_t=\log P_t$; raw returns are inappropriate because their lag-1
autocorrelation is $\approx 0$ by construction in an efficient market. For each
leverage/flow variable we take $x_t=\log(\cdot)$. The trailing (causal) mean is used
deliberately, so a rising signal could in principle be acted on before the transition.

\paragraph{Early-warning signals.}
Over a rolling window $w_r<w_d$ we compute two generic indicators of critical slowing
down \citep{scheffer2009early,dakos2012methods}: the variance $\operatorname{Var}(r)$
and the lag-1 autocorrelation $\rho_1=\operatorname{corr}(r_t,r_{t-1})$.

\paragraph{Trend test.}
We quantify ``rising toward the transition'' with the Kendall rank correlation
$\tau$ between each indicator and time over a pre-cascade window ending at the cascade
onset $t^\star$; $\tau>0$ with $p<0.05$ is a positive, significant early-warning trend.
The onset is detected identically across events as the minute ending the most
negative 60-minute log-return \emph{within the documented crash day} (unrestricted
detection inside a 2-month file can select a different, smaller crash); October
2025 is pinned to the documented 20:50 UTC collateral-depeg nucleation
\citep{coindesk2025usde,amberdata2025oct}.

\paragraph{Robustness.}
A single $(w_d,w_r,\text{pre-window})$ choice is fragile. We sweep a grid (detrend
$\sim$2--16\,h, roll $\sim$0.5--4\,h, pre-window $\in\{1,2,3\}$\,d; 39 valid configs
per variable) and report, per variable and indicator, the fraction of configs that are
positive and significant. A real signal survives most of the grid; a fluke does not.

\section{Results}
\label{sec:results}

\subsection{Crash anatomy}
\begin{figure}[t]
  \centering
  \includegraphics[width=0.8\linewidth]{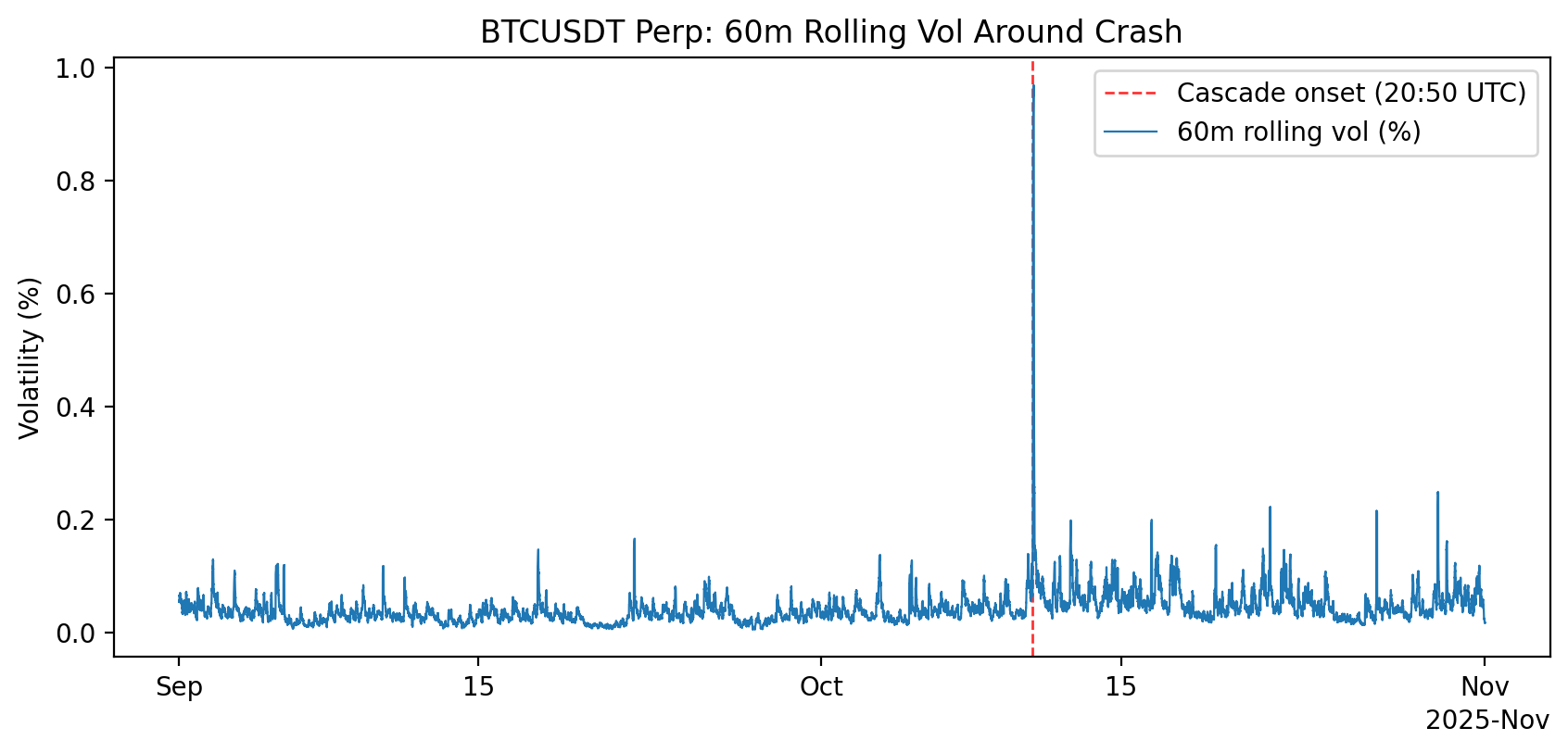}
  \caption{60-minute rolling volatility of BTCUSDT perp around the cascade. Source:
  EXP-000.}
  \label{fig:vol}
\end{figure}

\begin{figure}[t]
  \centering
  \includegraphics[width=0.8\linewidth]{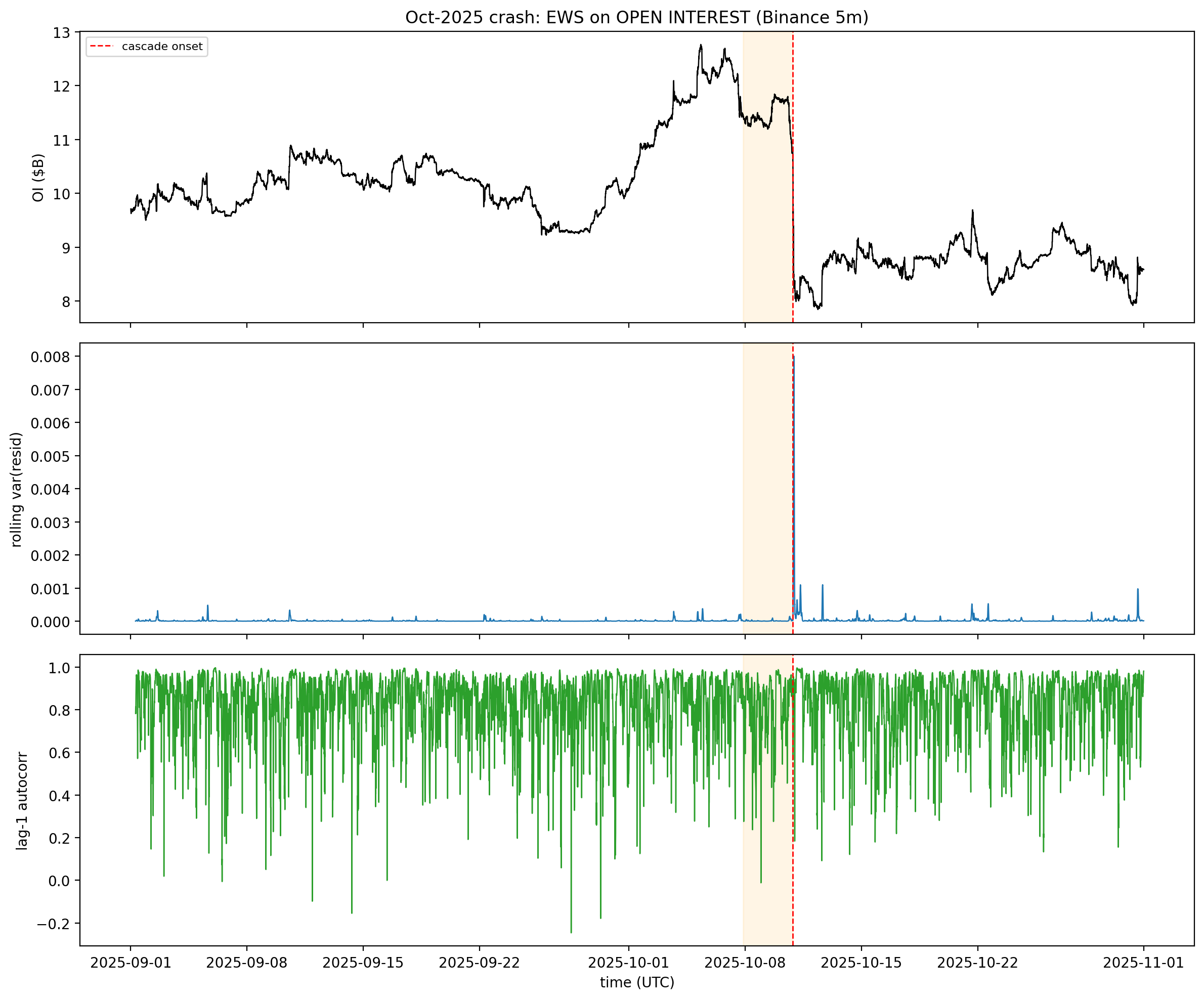}
  \caption{EWS on open interest (5-minute). Detrended-residual variance and lag-1
  autocorrelation into the cascade onset (red). Source: EXP-002.}
  \label{fig:ews_oi}
\end{figure}

Daily long-liquidation notionals over 2022--2025 have a Gini coefficient of
0.677; \textbf{10 October 2025 is the single largest long-liquidation day} in the
sample at \$1.05\,billion (versus $\sim$\$130\,million the
prior day). Intraday open interest falls $-24.6\%$ across the cascade
(Fig.~\ref{fig:ews_oi}); the coincident volatility burst is shown in
Fig.~\ref{fig:vol}.

\subsection{Price shows no early warning}
\begin{figure}[t]
  \centering
  \includegraphics[width=0.8\linewidth]{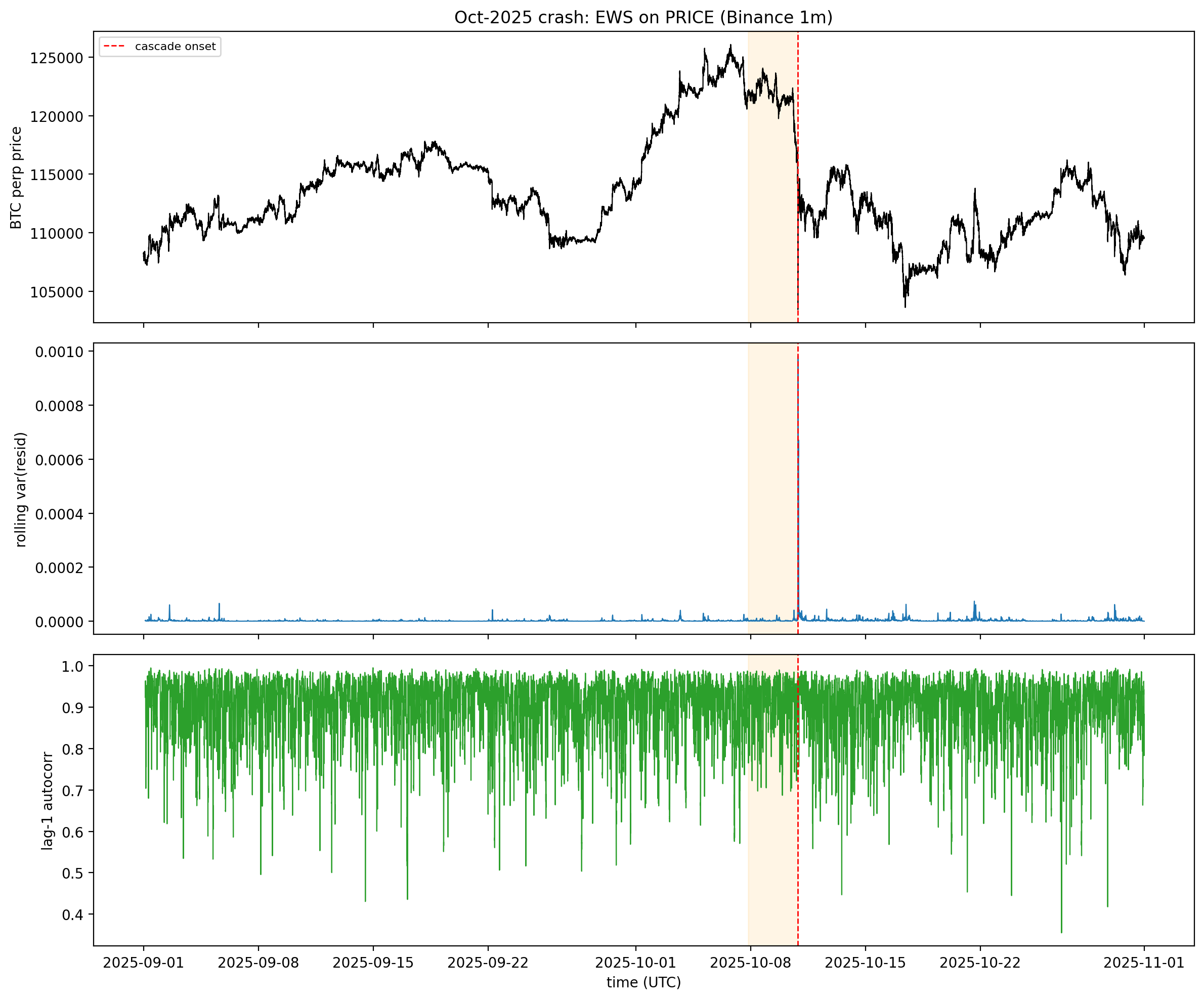}
  \caption{EWS on price (1-minute). The variance spike is coincident with the cascade,
  with no pre-cascade build-up; autocorrelation does not rise. Source: EXP-001.}
  \label{fig:ews_price}
\end{figure}

On the detrended log-price residual, lag-1 autocorrelation shows \emph{no} pre-cascade
rise: it is positive and significant in only $4/39$ configurations (median
$\tau=-0.038$). Rolling variance does trend up ($34/39$), but---as
Section~\ref{subsec:panel} shows---it does so for nearly every variable, so it does not
discriminate. The residual-variance spike is coincident with, not precursory to, the
cascade (Fig.~\ref{fig:ews_price}).

\subsection{The leverage/order-flow panel (October 2025, in-sample)}
\label{subsec:panel}
\begin{figure}[t]
  \centering
  \includegraphics[width=\linewidth]{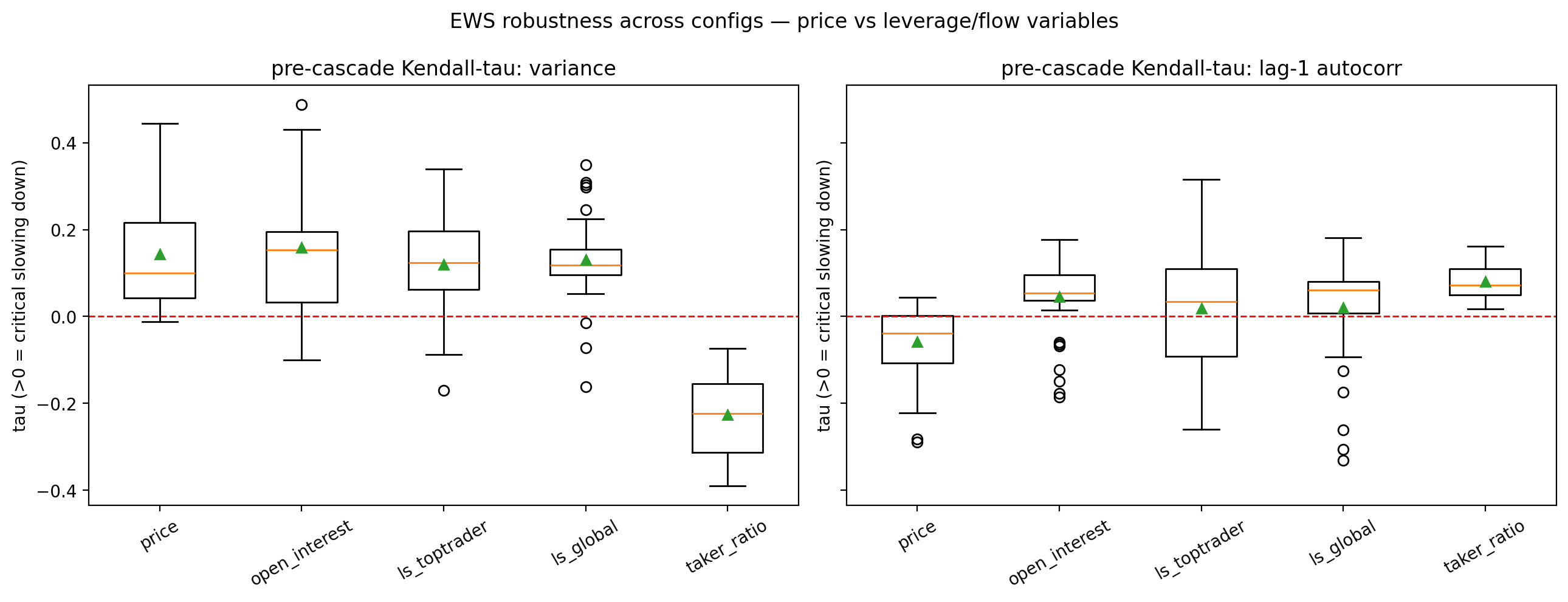}
  \caption{In-sample (October 2025) robustness across 39 configurations per variable.
  Right (lag-1 autocorrelation): price straddles/below zero while the leverage/flow
  variables sit above zero---the pattern that fails to replicate in Fig.~\ref{fig:oos}.
  Left (variance): non-discriminating. Source: EXP-004.}
  \label{fig:box}
\end{figure}

Table~\ref{tab:robust} and Fig.~\ref{fig:box} summarise the robustness sweep across
price and four leverage/flow variables \emph{for the October 2025 event}.

\begin{table}[t]
  \centering
  \caption{Pre-cascade Kendall-$\tau$ robustness across 39 configurations per
  variable. Cells show the count positive-and-significant ($p<0.05$) with the median
  $\tau$ in parentheses. Rising variance is non-specific; lag-1 autocorrelation
  discriminates price from the leverage/flow variables. Source: EXP-004.}
  \label{tab:robust}
  \begin{tabular}{lcc}
    \toprule
    Variable & Variance (pos\,\&\,sig) & Lag-1 AR (pos\,\&\,sig) \\
    \midrule
    Price                    & $34/39$ $(+0.101)$ & $4/39$  $(-0.038)$ \\
    Open interest            & $26/39$ $(+0.153)$ & $23/39$ $(+0.053)$ \\
    Top-trader L/S ratio     & $30/39$ $(+0.123)$ & $16/39$ $(+0.034)$ \\
    Global L/S account ratio & $35/39$ $(+0.119)$ & $21/39$ $(+0.060)$ \\
    Taker buy/sell ratio     & $0/39$  $(-0.223)$ & $29/39$ $(+0.072)$ \\
    \bottomrule
  \end{tabular}
\end{table}

Two facts stand out \emph{within this event}. First, \textbf{rising variance is
non-specific}: it is positive across most configurations for price and for the
positioning variables, and it \emph{falls} for taker flow---so on its own it is not
evidence of criticality. Second, \textbf{lag-1 autocorrelation appears to discriminate}:
it is essentially absent in price ($4/39$) but present across all four leverage/flow
variables ($16$--$29/39$), strongest in taker order flow ($29/39$) and open interest
($23/39$). Taken alone, this would suggest the critical-slowing-down signature lives in
the leverage structure rather than in price. Section~\ref{subsec:oos} shows this
inference does not survive out of sample.

\subsection{Out-of-sample: the pattern does not replicate}
\label{subsec:oos}
\begin{figure}[t]
  \centering
  \includegraphics[width=\linewidth]{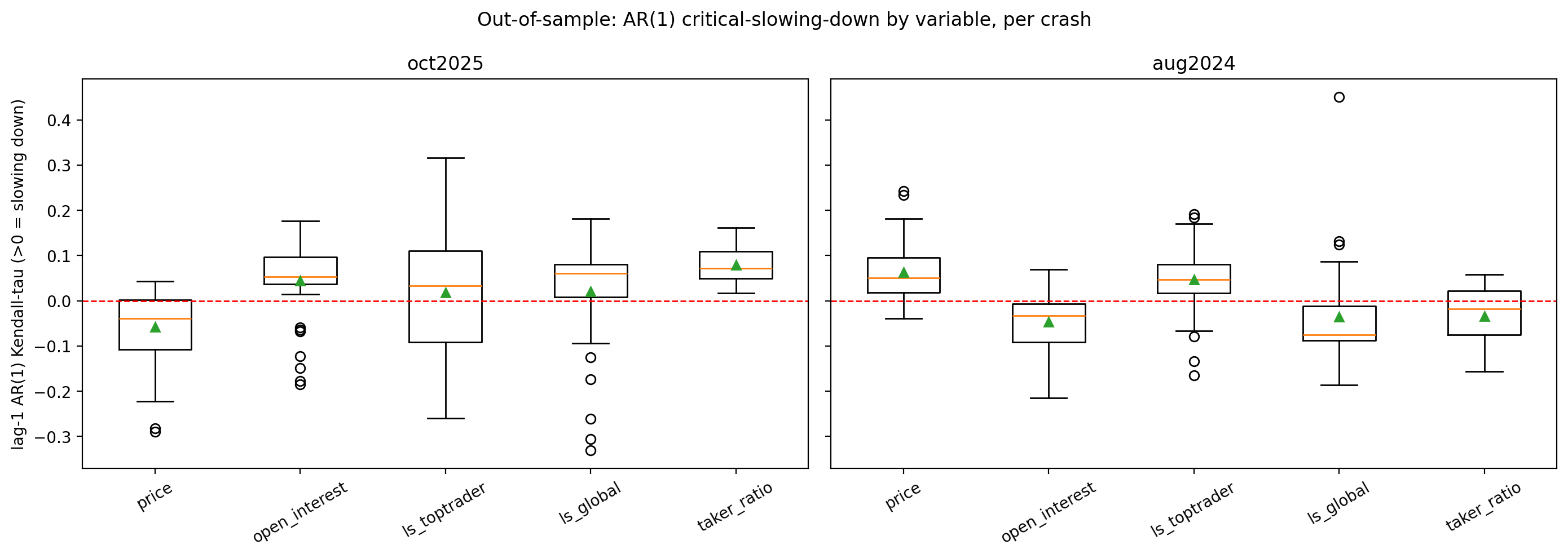}
  \caption{Out-of-sample comparison. Lag-1 autocorrelation Kendall-$\tau$ by variable
  for each crash. The sign pattern across variables inverts between events: no
  event-invariant early-warning variable. Source: EXP-005.}
  \label{fig:oos}
\end{figure}

We repeat the identical analysis on the August 2024 cascade. The October 2025 pattern
\textbf{does not replicate---it inverts} (Table~\ref{tab:oos}, Fig.~\ref{fig:oos}). In
August 2024 it is \emph{price} that carries the lag-1 autocorrelation signature
($28/39$), while open interest, the global long/short ratio and taker flow do
\emph{not} ($3/39$, $4/39$, $1/39$). The only variable positive in both events is the
top-trader long/short ratio ($16/39$ then $17/39$), and only weakly. Rolling variance
remains non-specific in both events, with taker-flow variance falling in both---the sole
consistent feature. Across the two crashes there is \textbf{no event-invariant
early-warning variable}.

\begin{table}[t]
  \centering
  \caption{Out-of-sample comparison: lag-1 autocorrelation, count
  positive-and-significant of 39 configs (median $\tau$). The variable carrying the
  signature differs between events; only top-trader L/S is (weakly) positive in both.
  Source: EXP-005.}
  \label{tab:oos}
  \begin{tabular}{lcc}
    \toprule
    Variable & Oct 2025 & Aug 2024 \\
    \midrule
    Price                    & $4/39$  $(-0.038)$ & $28/39$ $(+0.050)$ \\
    Open interest            & $23/39$ $(+0.053)$ & $3/39$  $(-0.033)$ \\
    Top-trader L/S ratio     & $16/39$ $(+0.034)$ & $17/39$ $(+0.047)$ \\
    Global L/S account ratio & $21/39$ $(+0.060)$ & $4/39$  $(-0.075)$ \\
    Taker buy/sell ratio     & $29/39$ $(+0.072)$ & $1/39$  $(-0.018)$ \\
    \bottomrule
  \end{tabular}
\end{table}

\subsection{The distribution of crash signatures ($n=7$)}
\label{subsec:distribution}
\begin{figure}[t]
  \centering
  \includegraphics[width=\linewidth]{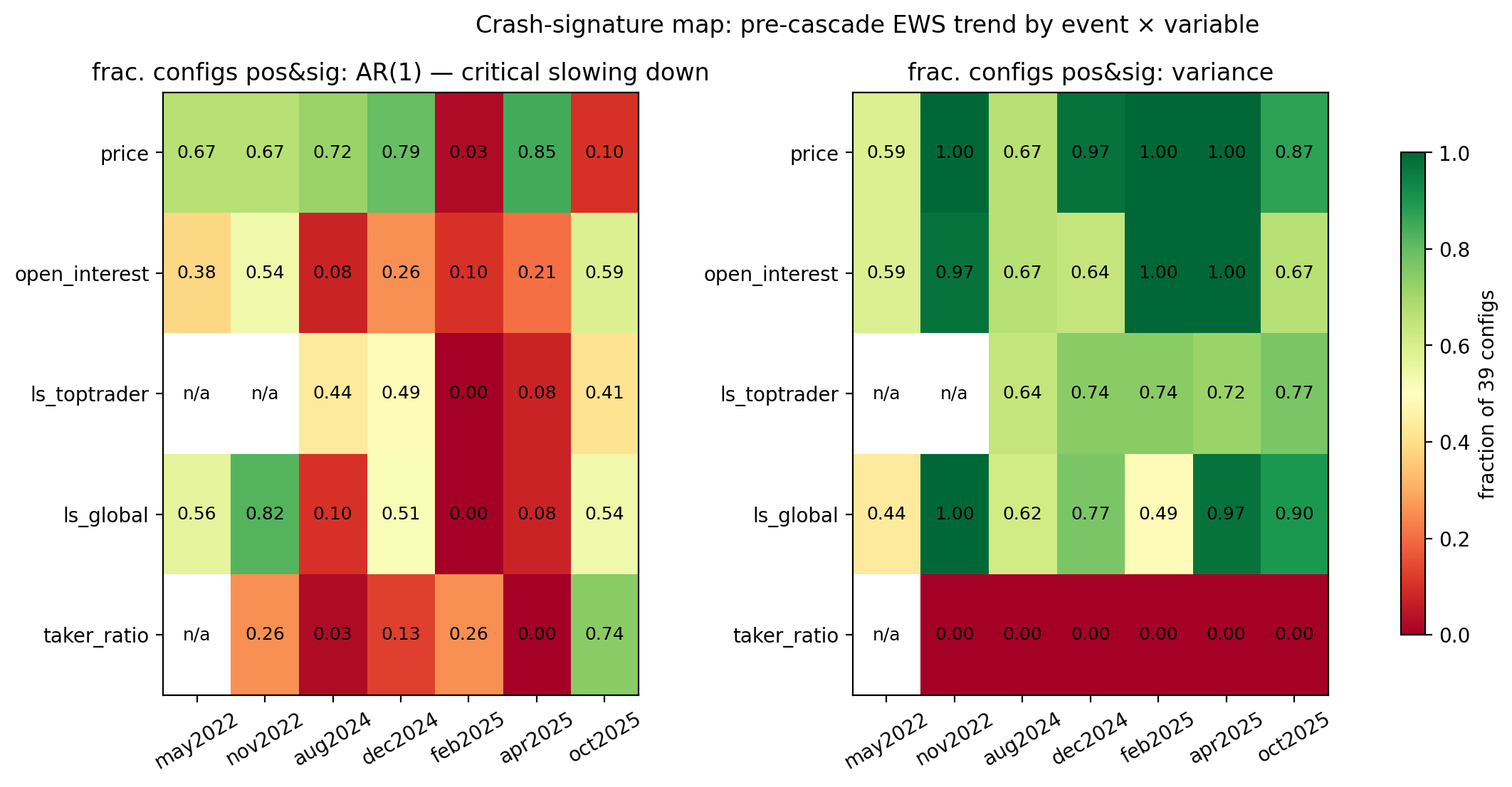}
  \caption{Crash-signature map: fraction of 39 configurations with a positive,
  significant pre-cascade trend, per event and variable (left: lag-1
  autocorrelation; right: variance). No variable is event-invariant; the two
  sudden-news events (Feb/Oct 2025) lack the price signature the other five
  carry; taker-ratio \emph{variance} falls everywhere. Source: EXP-006.}
  \label{fig:heatmap}
\end{figure}

Widening to all seven events (Table~\ref{tab:ar1_events};
Fig.~\ref{fig:heatmap}) settles the question the two-event comparison raised:
\begin{enumerate}
  \item \textbf{No variable is event-invariant.} The top-trader long/short
        ratio---the only candidate surviving the two-event comparison---fails in
        the 2025 tariff events ($0/39$ and $3/39$).
  \item \textbf{A two-type structure.} Price shows the critical-slowing-down
        signature in five of seven events ($0.67$--$0.85$ of configurations) and
        is absent in exactly two---October 2025 and February 2025, the two
        sudden-news (tariff) shocks. Anticipated or endogenous-buildup cascades
        carry a price signature; pure exogenous shocks carry none (February
        2025 shows no precursor in \emph{any} variable) or relocate it to
        leverage/flow (October 2025).
  \item \textbf{October 2025 is the outlier, not the rule} ($1/7$)---a caution
        for single-event criticality claims: the event that suggests the
        ``criticality lives in leverage'' hypothesis is the one event where it
        holds.
\end{enumerate}

\begin{table}[t]
  \centering
  \caption{Lag-1 autocorrelation: fraction of 39 configurations
  positive-and-significant, per event (chronological). n/a: series not yet
  published in the 2022 dumps. Source: EXP-006.}
  \label{tab:ar1_events}
  \begin{tabular}{lccccccc}
    \toprule
    Variable & May22 & Nov22 & Aug24 & Dec24 & Feb25 & Apr25 & Oct25 \\
    \midrule
    Price          & .67 & .67 & .72 & .79 & \textbf{.03} & .85 & \textbf{.10} \\
    Open interest  & .38 & .54 & .08 & .26 & .10 & .21 & .59 \\
    Top-trader L/S & n/a & n/a & .44 & .49 & \textbf{.00} & .08 & .41 \\
    Global L/S     & .56 & .82 & .10 & .51 & .00 & .08 & .54 \\
    Taker ratio    & n/a & .26 & .03 & .13 & .26 & .00 & .74 \\
    \bottomrule
  \end{tabular}
\end{table}

\subsection{The one invariant, placebo-tested: order-flow variance compression}
\label{subsec:placebo}
\begin{figure}[t]
  \centering
  \includegraphics[width=0.75\linewidth]{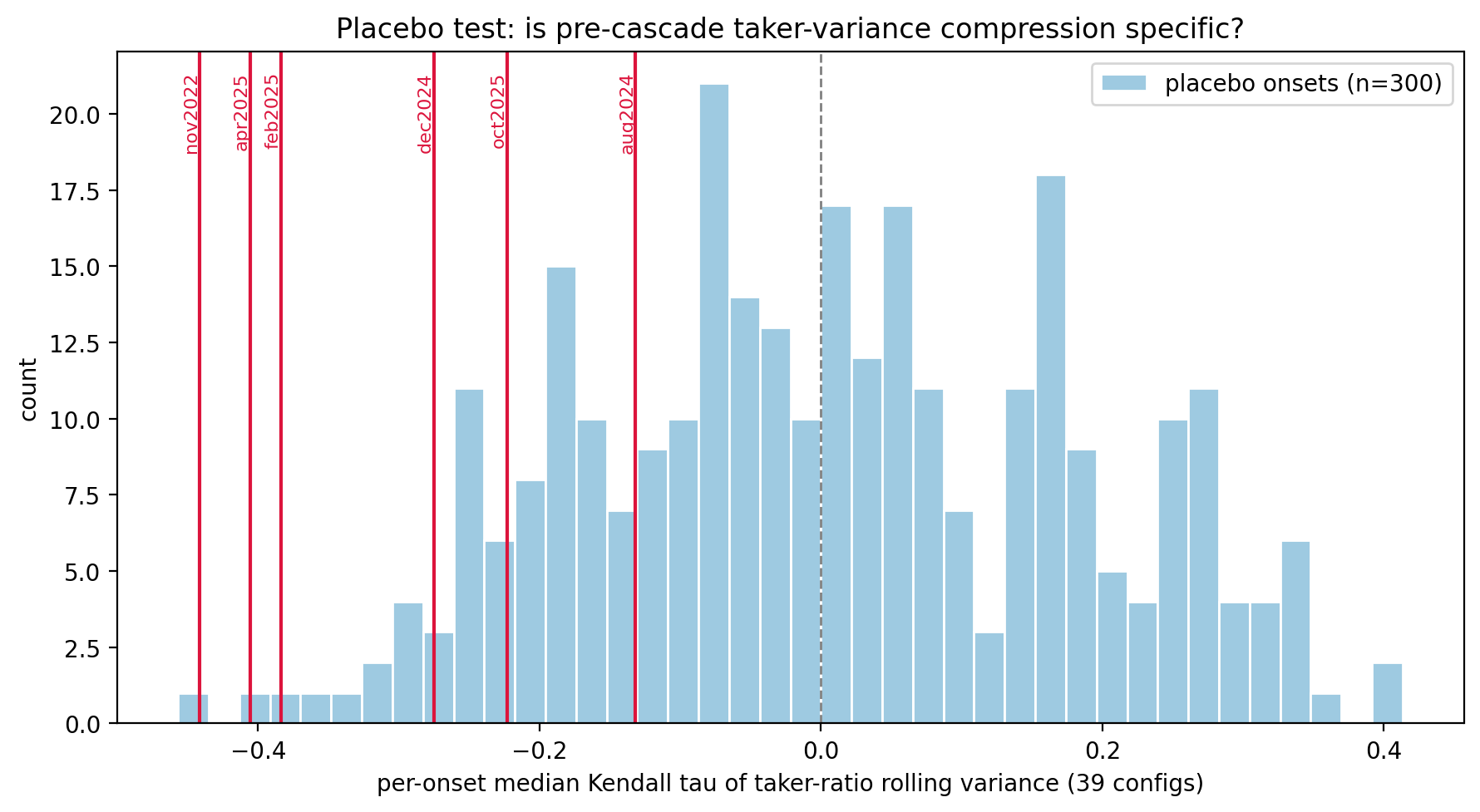}
  \caption{Placebo test for taker-flow variance compression: per-onset median
  Kendall-$\tau$ over 300 ordinary-market onsets (histogram) vs the six real
  cascades (red lines). Source: EXP-007.}
  \label{fig:placebo}
\end{figure}

One regularity survives every event with available data: the variance of the
taker buy/sell ratio \emph{falls} before all six cascades (median $\tau$
$-0.13$ to $-0.44$)---opposite in sign to the textbook variance-rising signal.
Because a mean-reverting ratio might show declining detrended variance in calm
regimes anyway, we test against 300 placebo onsets drawn from ordinary-market
windows (full pre-window available; no overlap with a studied onset; not
followed within 24\,h by a $>4\%$ drawdown, which would make the window the
pre-period of a different, unlabelled crash). The null is centred at zero
($P(\text{median}\,\tau<0)=0.49$): mean reversion does not explain the
compression. All six events fall in the null's left tail (four below the 5th
percentile; sign test $p\approx 0.014$; Fisher-combined
$p\approx 5\times10^{-6}$) (Fig.~\ref{fig:placebo}). The honest reading:
aggressive order flow going persistently one-sided is a \emph{population-level}
precursor---real, but too weak for per-event warning (two of six events overlap
the null individually).

\FloatBarrier
\section{Discussion}
\label{sec:discussion}

The central finding is negative and, we argue, more informative than the in-sample
story it overturns. Four threads develop it.

\paragraph{Heterogeneous crash signatures, with structure.}
The state variable carrying the critical-slowing-down signature is not shared
across cascades. At $n=7$ the heterogeneity acquires a candidate structure: the
five endogenous-buildup events carry a price signature; the two sudden-news
shocks do not---February 2025 is silent in every variable, October 2025
relocates the signature to leverage and flow. A mechanism suggests itself.
Critical slowing down is an estimator of how quickly perturbations relax, and it
can only be measured over an interval in which the system actually approaches
the instability. The five events with a price signature developed over hours to
days, as an insolvency, a funding stress or a policy date was progressively
absorbed; the pre-window contains an approach, and the price residual has time to
acquire the memory the theory predicts. The two exceptions were repriced within
minutes of an unscheduled announcement. When the shock arrives on a timescale
shorter than the system's relaxation time, there is no approach inside the
pre-window to detect, and an autocorrelation estimator has nothing to find. We
stress that this typology is a hypothesis \emph{generated} at $n=7$, not one
tested at $n=7$: with two events in the exceptional class it could not be
otherwise, and it must be checked on cascades we have not seen.

\paragraph{Variance is an unreliable early-warning signal here.}
Across the panel, rolling variance trends up for price and for the positioning
variables in essentially every event, irrespective of which variable ends up
carrying the autocorrelation signature and irrespective of crash type. A
statistic that rises before every cascade and also throughout ordinary market
regimes carries little discriminating information. Only its \emph{fall}, in
taker order flow, is consistent---and it is consistent in sign as well as in
presence. This aligns with the standing critique that variance-based early
warnings are prone to false positives, and it cautions specifically against the
variance-only claims that are easy to make in crypto markets, where volatility
clustering guarantees that some variance measure is rising almost always.

\paragraph{Perhaps the transition is not critical.}
There is a reading under which these results are not a measurement failure but a
prediction. Critical slowing down is a property of a system approaching a fold:
the approach must be slow relative to the observation window, and the system must
in fact be near the bifurcation. If a liquidation cascade is instead a
discontinuous, shock-driven transition---a supercritical branch entered when a
finite perturbation meets a sufficiently loaded book---then there is no slow
approach, and the absence of critical slowing down in the price is precisely what
the mechanism predicts rather than a defect of the indicator.
\citet{guttal2016lack} draw the analogous conclusion for equity meltdowns.
Nothing in this paper settles the question, because it cannot be settled from
single-series indicators: distinguishing a critical transition from a
discontinuous one is a statement about the \emph{order} of the transition, and
that is visible in the collective structure of the market---how its many
correlated components reorganise---rather than in any one variable at a time. We
take this up in a companion paper.

\paragraph{Implication for single-event studies, and for liquidity provision.}
A signature identified in one crash should not be assumed to generalise. The
October 2025 pattern was clean, mechanistically plausible, and wrong: had we
stopped at the event that motivated the study, we would have published the claim
that criticality lives in the leverage structure. Out-of-sample testing is not a
robustness appendix in this literature; it is the load-bearing step. For
liquidity provision the corollary is uncomfortable rather than actionable. A
market maker whose quoting model conditions on the local predictability of price
meets its worst conditions exactly when that predictability collapses, and our
results say the collapse is not reliably announced beforehand by the price series
itself. The one signal that does carry population-level information---aggressive
order flow going persistently one-sided---is observable in real time, and is
notably a \emph{flow} variable rather than a price variable. We offer this as
motivation for future work, not as a result: nothing established here shows that
any particular quoting rule would have helped.

\paragraph{Limitations.}
Seven events, a single venue, and moderate effect sizes throughout. Detrending
and window choices matter, and the robustness sweep mitigates rather than
removes that dependence. Funding settles on a fixed eight-hour cycle, and this
institutional periodicity can in principle imprint on the autocorrelation of
the flow and positioning residuals; the sweep spans detrending windows on both
sides of that period, but we do not explicitly deseasonalise it. The 2022 metrics dumps are incomplete: the top-trader
long/short ratio is unavailable for both 2022 events, and the taker ratio
additionally for May 2022, so the compression result rests on the six events
with a usable taker series. The placebo null is drawn from those same six
two-month files, which bounds its regime coverage to those periods and makes the
independence assumed by the joint $p$-values approximate; onsets within a file
share a regime. The null is conservative in one respect---unlabelled smaller
selloffs that survive the $4\%$ drawdown filter inflate compression under the
null---and this works against us, not for us. Finally, and most importantly, the
leverage variables are proxies for a liquidation state we cannot observe
(Section~\ref{sec:data}). These bound the claim to exactly what the data support.

\section{Conclusion and outlook}
\label{sec:conclusion}

A lag-1-autocorrelation early-warning pattern that looked clean within the
October 2025 cascade---present in leverage/order-flow, absent in price---fails
to replicate out of sample, and across seven cascades no variable is
event-invariant: the signature moves with the crash type (endogenous-buildup
vs.\ sudden-shock), and rolling variance is uniformly unreliable. The one
placebo-tested regularity, compression of taker order-flow variance, is a
population-level precursor rather than a per-event alarm.

Two conclusions follow. The first is cautionary: single-venue, single-variable
early-warning claims in crypto perpetual markets are fragile \emph{by
construction}. A venue is one node of an arbitrated multi-venue system, and the
post-mortems of October 2025 document contagion propagating across venues
through arbitrage \citep{amberdata2025oct}; a state variable measured on one node
need not be a state variable of the system.

The second is more consequential, and it is what we think this paper contributes
beyond a replication of \citet{guttal2016lack} in a new asset class. The
failures are not randomly distributed. Critical slowing down is absent exactly
where the destabilising mechanism is most violent and most abrupt, and present
where the market had time to absorb its own bad news. That is the pattern one
expects if these cascades are not critical transitions at all, but discontinuous
transitions driven by finite shocks meeting a loaded book. Two continuations
follow, and we pursue both in a companion paper: characterising the transition
through the collective correlation structure of the market, where its
\emph{order} is visible in a way that no single series can reveal; and asking
whether a state variable derived from the mechanism itself---the density of
liquidation thresholds near the price, weighted by the impact of the forced flow
that crosses them---succeeds where statistical proxies fail. Recovery of the
intraday liquidation microstructure discussed in Section~\ref{sec:data} is the
data prerequisite for the second, and the reason the question remains open here.

\FloatBarrier
\section*{Reproducibility}
All figures and numbers are regenerated by the scripts in the repository; each
quantitative claim cites a frozen experiment record (\texttt{notes/experiments.md},
EXP-000--007).

\section*{Acknowledgements}
This work received no specific grant from any funding agency. It uses only
publicly available exchange data; no proprietary or paid data source was used.

\bibliographystyle{plainnat}
\bibliography{refs}

\end{document}